\documentclass[prl,10pt,aps,twocolumn,superscriptaddress,altaffilletter,nofootinbib]{revtex4-1}

\usepackage{graphicx}  % needed for figures
\usepackage{dcolumn}   % needed for some tables
\usepackage{bm}        % for math
\usepackage{amssymb}   % for math
\usepackage{epstopdf}
\usepackage{hyperref}
\usepackage{color}

\newcommand{\Bilgi}{Istanbul Bilgi University, High Energy Physics Research Center, Eyup, Istanbul, 34060,
Turkey}
\newcommand{\Saclay}{IRFU, CEA, Universit\'{e} Paris-Saclay, Gif-sur-Yvette, France}
\newcommand{\CERN}{European Organization for Nuclear Research (CERN), Gen\`eve, Switzerland}
\newcommand{\INR}{Institute for Nuclear Research (INR), Russian Academy of Sciences, Moscow, Russia}
\newcommand{\MPE}{Max-Planck-Institut f\"{u}r Extraterrestrische Physik, Garching, Germany}
\newcommand{\Trieste}{Istituto Nazionale di Fisica Nucleare (INFN), Sezione di Trieste and Universit\`a di
Trieste,Trieste, Italy}
\newcommand{\Zaragoza}{Grupo de Investigaci\'{o}n de F\'{\i}sica Nuclear y Astropart\'{\i}culas, Universidad de
Zaragoza,Zaragoza, Spain }
\newcommand{\Columbia}{Physics Department and Columbia Astrophysics Laboratory, Columbia University, New York,
NY 10027, USA}
\newcommand{\DTU}{DTU Space, National Space Institute, Technical University of Denmark, Lyngby, Denmark}
\newcommand{\Chicago}{Enrico Fermi Institute and KICP, University of Chicago, Chicago, IL 60637, USA}
\newcommand{\Thessaloniki}{Aristotle University of Thessaloniki, Thessaloniki, Greece}
\newcommand{\Demokritos}{National Center for Scientific Research ``Demokritos'', Athens, Greece}
\newcommand{\Freiburg}{Albert-Ludwigs-Universit\"{a}t Freiburg, Freiburg, Germany}
\newcommand{\Patras}{Physics Department, University of Patras, Patras, Greece}
\newcommand{\Athens}{National Technical University of Athens, Athens, Greece}

\newcommand{\Vancouver}{Department of Physics and Astronomy, University of British Columbia, Vancouver, Canada}
\newcommand{\Darmstadt}{Technische Universit\"{a}t Darmstadt, IKP, Darmstadt, Germany}
\newcommand{\Frankfurt}{Johann Wolfgang Goethe-Universit\"at, Institut f\"ur Angewandte Physik, Frankfurt am
Main, Germany}
\newcommand{\Zagreb}{Rudjer Bo\v{s}kovi\'{c} Institute, Zagreb, Croatia}
\newcommand{\MPP}{Max-Planck-Institut f\"{u}r Physik (Werner-Heisenberg-Institut), M\"unchen, Germany}
\newcommand{\LLNL}{Lawrence Livermore National Laboratory, Livermore, CA 94550, USA}
\newcommand{\MPIS}{Max-Planck-Institut f\"{u}r Sonnensystemforschung, G\"{o}ttingen, Germany}

\newcommand{\Korea}{School of Space Research, Kyung Hee University, Yongin, Republic of Korea.}

\newcommand{\Rijeka}{Department of Physics and Centre for Micro and Nano Sciences and Technologies, University
of Rijeka, Rijeka, Croatia.}

\newcommand{\Camerino}{Physics Division, School of  Sciences and Technology, University of Camerino, Camerino,
Italy.}
\newcommand{\California}{Dept.\ of Physics and Astronomy, University of California, Irvine, CA 92697, USA.}
\newcommand{\Bonn}{Physikalisches Institut, University of Bonn, Germany.}
\newcommand{\CAPP}{Center for Axion and Precision Physics Research, Institute for Basic Science (IBS), Daejeon
34141, Republic of Korea.}
\newcommand{\KAIST}{Department of Physics, Korea Advanced Institute of Science and Technology (KAIST), Daejeon
34141, Republic of Korea}
\newcommand{\ELI}{Extreme Light Infrastructure - Nuclear Physics (ELI-NP), 077125 Magurele, Romania}
\newcommand{\IHEP}{Institute of High Energy Physics, Chinese Academy of Sciences, Beijing}
\newcommand{\CERijeka}{ Photonics and Quantum Optics Unit, Center of Excellence for Advanced Materials and Sensing Devices, University of Rijeka, Rijeka, Croatia }

\newcommand{\remove}[1]{}
\newcommand{\IGI}[1]{#1}
\newcommand{\GGR}[1]{#1}

\begin{document}

\title{%\IGI{CAST solar axion search achieves forefront sensitivity}}
New CAST Limit on the Axion--Photon Interaction}

\author{    V.~Anastassopoulos}\affiliation{\Patras}
\author{    S.~Aune  }\affiliation{\Saclay}
\author{    K.~Barth  }\affiliation{\CERN}
\author{    A.~Belov  }\affiliation{\INR}
\author{    H.~Br\"auninger  }\affiliation{\MPE}
\author{    G.~Cantatore  }\affiliation{\Trieste}
\author{    J.~M.~Carmona  }\affiliation{\Zaragoza}
\author{    J.~F.~Castel  }\affiliation{\Zaragoza}
\author{    S.~A.~Cetin  }\affiliation{\Bilgi}
\author{    F.~Christensen }\affiliation{\DTU}
\author{    J.~I.~Collar  }\affiliation{\Chicago}
\author{    T.~Dafni  }\affiliation{\Zaragoza}
\author{    M.~Davenport  }\affiliation{\CERN}
\author{    T.~Decker  }\affiliation{\LLNL}
\author{    A.~Dermenev  }\affiliation{\INR}
\author{	K.~Desch}\affiliation{\Bonn}
\author{    C.~Eleftheriadis  }\affiliation{\Thessaloniki}
\author{    G.~Fanourakis  }\affiliation{\Demokritos}
\author{    E.~Ferrer-Ribas  }\affiliation{\Saclay}
\author{    H.~Fischer  }\affiliation{\Freiburg}
\author{    J.~A.~Garc\' ia  }\altaffiliation[Present address: ]{\IHEP}\affiliation{\Zaragoza}
\author{    A.~Gardikiotis  }\affiliation{\Patras}
\author{    J.~G.~Garza  }\affiliation{\Zaragoza}
\author{    E.~N.~Gazis  }\affiliation{\Athens}
\author{    T.~Geralis  }\affiliation{\Demokritos}
\author{    I.~Giomataris  }\affiliation{\Saclay}
\author{    S.~Gninenko  }\affiliation{\INR}
\author{    C.~J.~Hailey  }\affiliation{\Columbia}
\author{    M.~D.~Hasinoff  }\affiliation{\Vancouver}
\author{    D.~H.~H.~Hoffmann  }\affiliation{\Darmstadt}
\author{    F.~J.~Iguaz  }\affiliation{\Zaragoza}
\author{    I.~G.~Irastorza  }\email[Corresponding author: ]{igor.irastorza@cern.ch}\affiliation{\Zaragoza}
\author{    A.~Jakobsen  }\affiliation{\DTU}
\author{    J.~Jacoby  }\affiliation{\Frankfurt}
\author{    K.~Jakov\v ci\' c  }\affiliation{\Zagreb}
\author{ 	J. Kaminski}\affiliation{\Bonn}
\author{    M.~Karuza  }\altaffiliation[Also at: ]{\CERijeka}\affiliation{\Trieste}\affiliation{\Rijeka}
\author{    N.~Kralj  }\altaffiliation[Present addr.: ]{\Camerino}\affiliation{\Rijeka}
\author{    M.~Kr\v{c}mar  }\affiliation{\Zagreb}
\author{    S.~Kostoglou}\affiliation{\CERN}
\author{ 	Ch. Krieger}\affiliation{\Bonn}
\author{    B.~Laki\'{c}  }\affiliation{\Zagreb}
\author{    J.~M.~Laurent  }\affiliation{\CERN}
\author{    A.~Liolios  }\affiliation{\Thessaloniki}
\author{    A.~Ljubi\v{c}i\'{c}  }\affiliation{\Zagreb}
\author{    G.~Luz\'on  }\affiliation{\Zaragoza}
\author{    M.~Maroudas}\affiliation{\Patras}
\author{    L.~Miceli  }\affiliation{\CAPP}
\author{    S.~Neff  }\affiliation{\Darmstadt}
\author{    I.~Ortega  }\affiliation{\Zaragoza}\affiliation{\CERN}
\author{    T.~Papaevangelou  }\affiliation{\Saclay}
\author{    K.~Paraschou  }\affiliation{\Thessaloniki}
\author{    M.~J.~Pivovaroff  }\affiliation{\LLNL}
\author{    G.~Raffelt  }\affiliation{\MPP}
\author{    M.~Rosu  }\altaffiliation[Present addr.: ]{\ELI}\affiliation{\Darmstadt}
\author{    J.~Ruz  }\affiliation{\LLNL}
\author{    E.~Ruiz Ch\'oliz  }\affiliation{\Zaragoza}
\author{    I.~Savvidis  }\affiliation{\Thessaloniki}
\author{    S.~Schmidt  }\affiliation{\Bonn}
\author{    Y.~K.~Semertzidis  }\altaffiliation[Also at.: ]{\KAIST}\affiliation{\CAPP}
\author{    S.~K.~Solanki  }\altaffiliation[Also at: ]{\Korea}\affiliation{\MPIS}
\author{    L.~Stewart  }\affiliation{\CERN}
\author{    T.~Vafeiadis  }\affiliation{\CERN}
\author{    J.~K.~Vogel  }\affiliation{\LLNL}
\author{    S.~C.~Yildiz  }\altaffiliation[Present addr.: ]{\California}\affiliation{\Bilgi}
\author{    K.~Zioutas  }\affiliation{\Patras}\affiliation{\CERN}

\collaboration{CAST Collaboration} \noaffiliation

\date{\today}

\begin{abstract}
\GGR{New low-mass particles, notably axions $a$, are motivated by
compelling theoretical ideas and the dark matter of the universe.
Such particles would also emerge abundantly from the hot interior of stars.
The CERN Axion Solar Telescope (CAST) searches for x-rays that
are expected after $a\to\gamma$ conversion in the 9~T magnetic field of a refurbished
LHC test magnet ($L=9.26$~m) that can be directed toward the Sun.
The absence of a significant signal above background in the 2013--2015 data
provides a world leading
limit of $g_{a\gamma}<0.66\times10^{-10}~{\rm GeV}^{-1}$ (95\% C.L.)
on the axion-photon coupling strength, similar to the most restrictive
astrophysical bounds. The magnet pipes were evacuated to avoid
x-ray refraction, implying this bound applies for $m_a\lesssim0.02$~eV.
Compared with the first vacuum phase (2003--2004),
the signal-to-noise was increased by about a factor of 3 using
low-background detectors and a new x-ray telescope.
These innovations also serve as pathfinders for a possible
next-generation axion helioscope that is needed if parameters
beyond the CAST benchmark are to be explored.}
\end{abstract}

\maketitle

{\ }
\newpage
{\ }

\begin{figure*}
\includegraphics[width=1.0\textwidth]{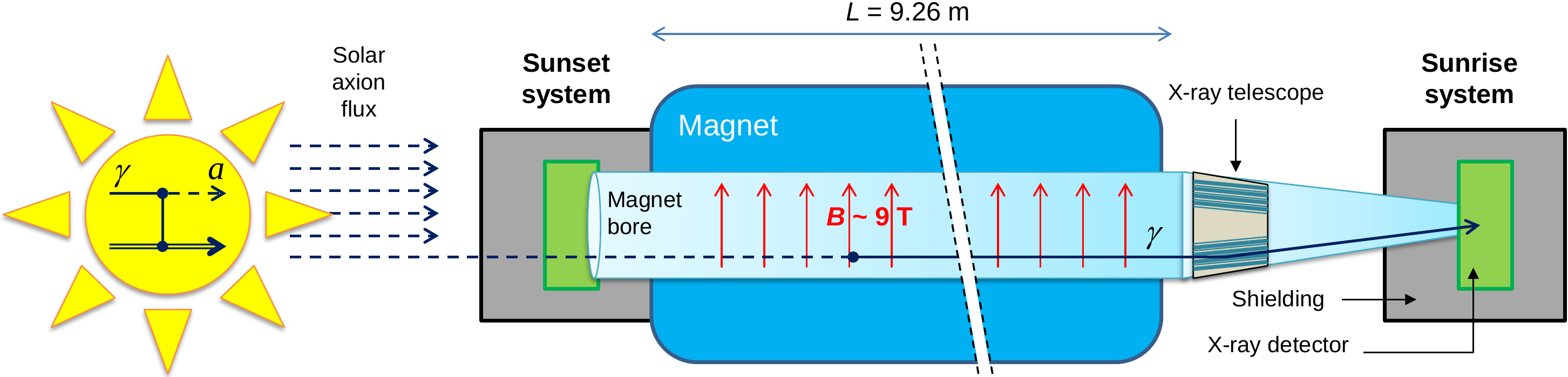}
\caption{\GGR{Sketch of the CAST helioscope at CERN to search for solar axions. These
hypothetical low-mass bosons are produced in the Sun by Primakoff scattering on charged particles
and converted back to x-rays in the $B$-field of an LHC test magnet. The two straight conversion pipes have
a cross section of 14.5~cm$^2$ each. The magnet can move by $\pm8^\circ$ vertically
and $\pm40^\circ$ horizontally, enough
to follow the Sun for about 1.5~h at dawn and dusk with opposite ends.
Separate detection systems can search for axions at sunrise and
sunset, respectively. The sunrise system is equipped with an x-ray telescope (XRT) to
focus the signal on a small detector area, strongly increasing signal-to-noise.
Our new results were achieved thanks to an XRT specifically built for CAST and
improved low-noise x-ray detectors.}
}\label{fig:cast}
\end{figure*}

\newpage

\section{Introduction}

Advancing the low-energy frontier is a key endeavor in the world-wide quest
for particle physics beyond the standard model and in the effort to identify
dark matter \cite{Olive:2016xmw,Jaeckel:2010ni}. Nearly massless pseudoscalar
bosons, often generically called axions, are particularly promising because
they appear in many extensions of the standard model. They can be dark matter
in the form of classical field oscillations that were excited in the early
universe, notably by the re-alignment mechanism \cite{AxionBook}. One
particularly well motivated case is the QCD axion, the eponym for all such
particles, which appears as a consequence of the Peccei-Quinn mechanism to
explain the absence of CP-violating effects in QCD \cite{AxionBook}.

Axions were often termed ``invisible'' because of their extremely feeble
interactions, yet they are the target of a fast-growing international
landscape of experiments. Numerous existing and foreseen projects assume that
axions are the galactic dark matter and use a variety of techniques that are
sensitive to different interaction channels and optimal in different mass
ranges \cite{AxDM-Searches,Patras}. Independently of the dark matter assumption, one
can search for new forces mediated by these low-mass bosons \cite{Forces} or
the back-reaction on spinning black holes (superradiance)~\cite{Superradiance}.
Stellar energy-loss arguments provide restrictive
limits that can guide experimental efforts and in some cases may even suggest
new loss channels \cite{AxionBook,Ayala:2014pea,Giannotti:2015kwo}.

The least model-dependent search strategies use the production and detection
of axions and similar particles by their generic two-photon coupling. It is
given by the vertex \smash{${\cal L}_{a\gamma} =-\frac{1}{4}\,g_{a\gamma}
F^{\mu\nu}\widetilde F_{\mu\nu}a =g_{a\gamma} {\bf E}\cdot{\bf B}\,a$}, where
$a$ is the axion field, $F$ the electromagnetic field-strength tensor, and
$g_{a\gamma}$ a coupling constant of dimension (energy)$^{-1}$.  Notice that
we use natural units with $\hbar=c=k_{\rm B}=1$. This vertex enables the
decay $a\to\gamma\gamma$, the Primakoff production in stars, i.e., the
$\gamma\to a$ scattering in the Coulomb fields of charged particles in the
stellar plasma, and the coherent conversion $a\leftrightarrow\gamma$ in
laboratory or astrophysical $B$-fields \cite{Sikivie:1983ip,Raffelt:1987im}.

\GGR{The helioscope concept, in particular, uses a dipole magnet directed at the
Sun to convert axions to x-rays (see Fig.~\ref{fig:cast} for a sketch).
Solar axions emerge from many thermal processes, depending on their model-dependent
interaction channels. We specifically consider axion production by Primakoff
scattering of thermal photons deep in the Sun, a process that depends on the
same coupling constant $g_{a\gamma}$ which is also used for detection.}

\begin{figure}
\includegraphics[width=1.0\columnwidth]{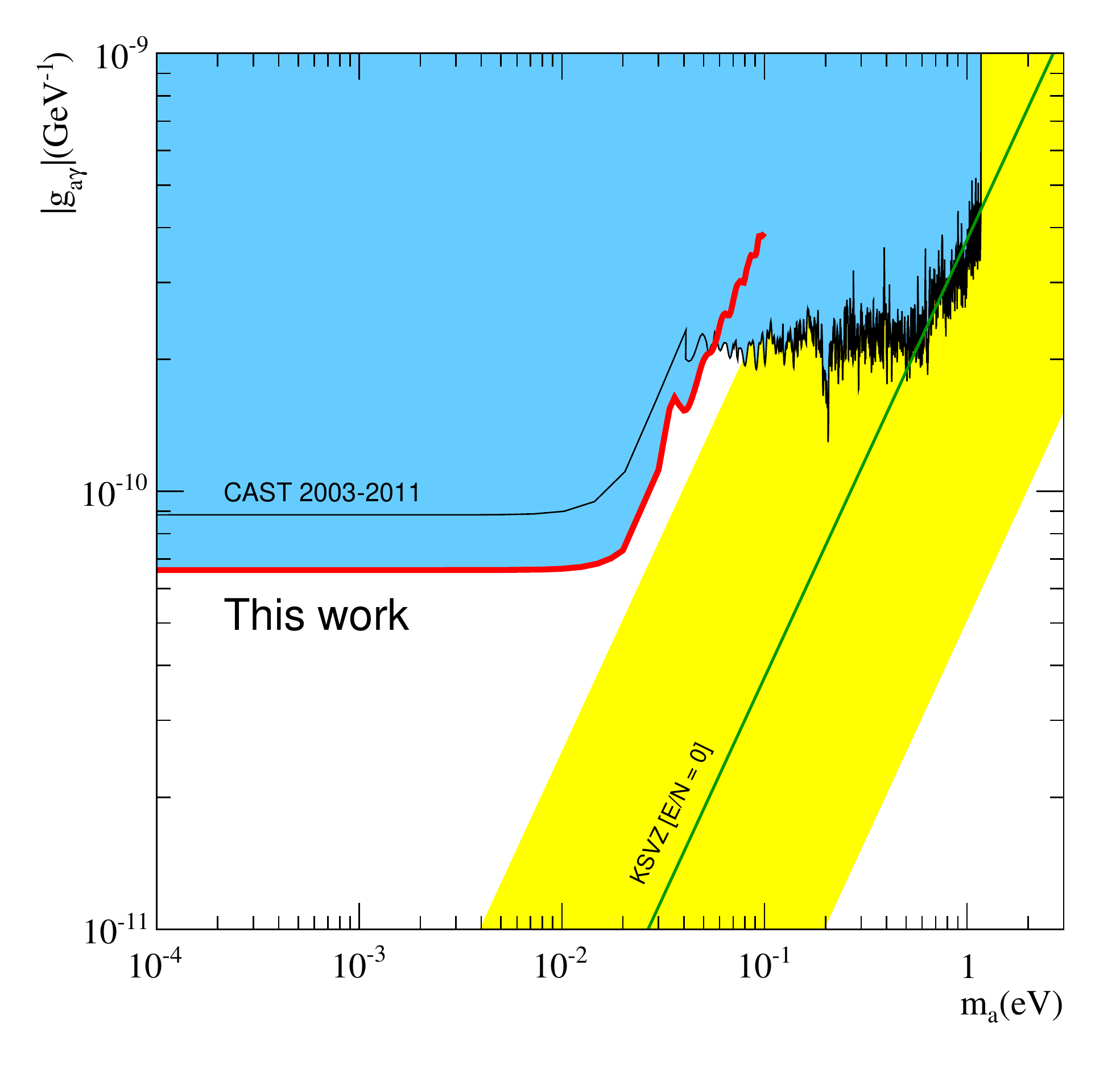}
\caption{CAST excluded region \IGI{(95\%~C.L.)} in the $m_a$--$g_{a\gamma}$--plane.
  {\em Solid black line:\/} Envelope of all  CAST results from 2003--2011
  data \cite{Zioutas:2004hi,Andriamonje:2007ew,Arik:2008mq, Arik:2011rx, Arik:2013nya, Arik:2015cjv}.
  {\em Solid red line:\/} Exclusion from the data here presented.
  {\em Diagonal yellow band:\/} Typical QCD axion models
  (upper and lower bounds set according to a prescription given in Ref.~\cite{DiLuzio:2016sbl}).
  {\em Diagonal green line:\/} The benchmark KSVZ axion model
  with $E/N=0$, where $g_{a\gamma}=(E/N-1.92)\,\alpha/(2\pi f_a)$ with $f_a$
  the axion decay constant.}\label{fig:limits}
\end{figure}

Since 2003, the CERN Axion Solar Telescope (CAST) has
explored the $m_a$--$g_{a\gamma}$ parameter space with this approach,
\GGR{more details to be given below}.
The black solid line in Fig.~\ref{fig:limits} is the envelope of all previous CAST results.
\GGR{The low-mass part $m_a\alt $0.02~eV corresponds to the first
phase 2003--2004 using evacuated magnet bores}
\cite{Zioutas:2004hi, Andriamonje:2007ew}.
The $a\to\gamma$ conversion probability in a homogeneous $B$ field over a
distance~$L$~is
\begin{equation}\label{eq:ConvProb}
P_{a\to\gamma}=\left(g_{a\gamma}B\,\frac{\sin(q L/2)}{q}\right)^2\,,
\end{equation}
where
$q=m_a^2/2E$ is the \hbox{$a$--$\gamma$} momentum transfer in vacuum. For
$L=9.26~{\rm m}$ and energies of a few keV, coherence is lost for
$m_a\agt0.02~{\rm eV}$, explaining the loss of sensitivity for larger $m_a$.

Later, CAST has explored this higher-mass range by filling the conversion
pipes with $^4$He \cite{Arik:2008mq, Arik:2015cjv}
and $^3$He \cite{Arik:2011rx,Arik:2013nya}
at variable pressure settings to
provide photons with a refractive mass and in this way match the $a$ and
$\gamma$ momenta. The sensitivity is smaller
because at each pressure setting, data were typically taken for a few hours
only. Despite this limitation, CAST has reached realistic QCD axion models
and has superseded previous solar axion searches using the helioscope
\cite{SUMICO} and Bragg
scattering technique \cite{Paschos:1993yf,Bernabei:2001ny}.
\GGR{(For a more complete list of previous solar axion constraints see Ref.~\cite{Olive:2016xmw}.)}
The CAST data were also interpreted in terms of other assumed axion
production channels in the Sun
\cite{Andriamonje:2009ar,Andriamonje:2009dx,Barth:2013sma}.
Moreover, CAST constraints
on other low-mass bosons include chameleons \cite{Anastassopoulos:2015yda}
and hidden photons \cite{Redondo:2015iea}.

\begin{figure*}[t!]
\centering
\includegraphics[width=\textwidth]{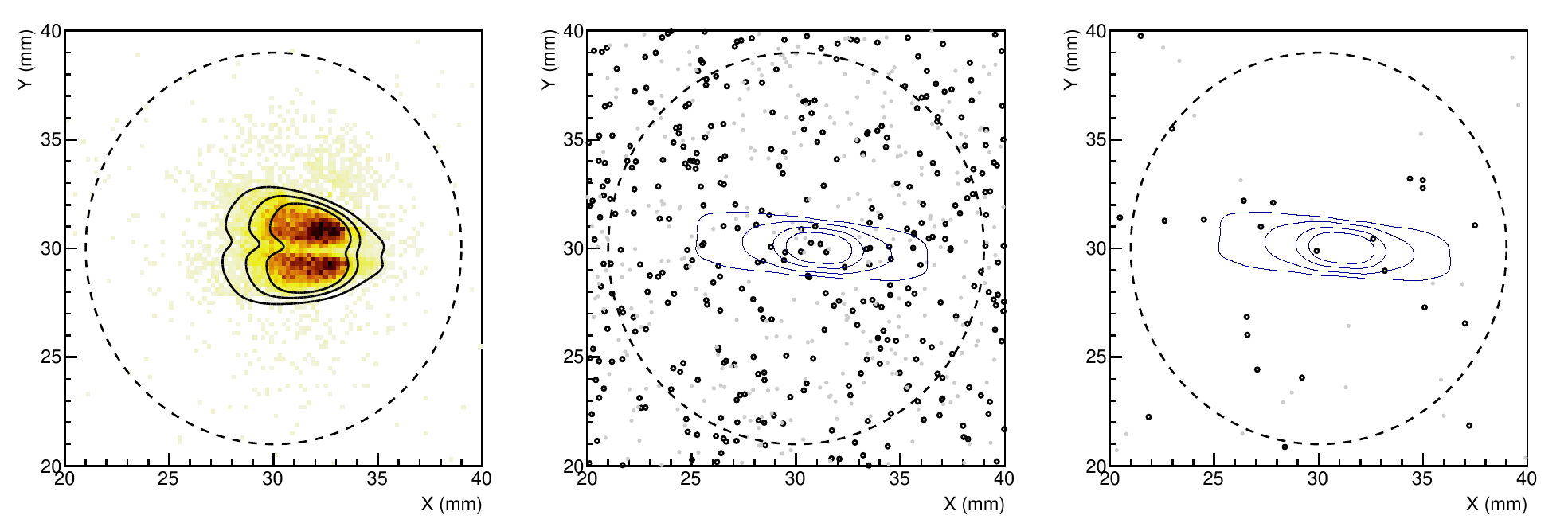}
\caption{\IGI{2D hitmap of events detected in the sunrise detector in a typical in-situ calibration run (left), as well as in the background (middle) and tracking (right) data (both K and L data sets \IGI{of Table~\ref{tab:datasets}}). The calibration is performed with an x-ray source placed $\sim$12~m away (at the sunset side of the magnet). The contours in the calibration run represent the 95\%, 85\% and 68\% signal-encircling regions from ray-trace simulations, taking into account the source size and distance. In the tracking and background plots, grey full circles represent events that pass all detector cuts but that are in coincidence with the muon vetoes, and therefore rejected. Black open circles represent final counts. Closed contours indicate the 99\%, 95\%, 85\% and 68\% signal-encircling regions out of detailed ray-trace simulations of the XRT plus spatial resolution of the detector. The large circle represents the region of detector exposed to daily energy calibration.}}
\label{fig:hitmaps}
\end{figure*}

During this long experimental program, CAST has used a variety of detection systems at both magnet ends,
including a multiwire time projection chamber~\cite{Autiero:2007uf}, several Micromegas
detectors~\cite{Abbon:2007ug}, a low-noise charged coupled device attached to a spare x-ray telescope (XRT)
from the ABRIXAS x-ray mission~\cite{Kuster:2007ue}, a $\gamma$-ray calorimeter~\cite{Andriamonje:2009ar}, and a silicon drift detector~\cite{Anastassopoulos:2015yda}.

In the latest data taking campaign (2013--2015), CAST has returned to evacuated
pipes, with an improvement in the sensitivity to solar axions \IGI{of about a factor $\sim$3 in signal-to-noise ratio} over a decade ago, thanks to the
development of novel detection systems, notably new Micromegas detectors with lower background levels, as well
as a new XRT built specifically for axion searches. These developments are also part of the activities to
define the detection technologies suitable for the proposed much larger next-generation axion helioscope IAXO
\cite{Armengaud:2014gea}.
We here report the results of this effort in CAST.

\section{Experiment and data taking}

CAST has utilized an LHC prototype dipole magnet~\cite{Zioutas:1998cc}
(magnetic field $B\sim9$~T, length $L=9.26$~m) with two parallel straight pipes
(cross-sectional area $S=2\times14.5\mbox{ cm}^{2}$). The magnet is mounted on a movable
platform ($\pm8^{\circ}$ vertical and $\pm40^{\circ}$ horizontal movement), allowing to follow the
Sun for about 1.5~h both at sunrise and sunset during the whole year. The pointing accuracy of the system is monitored to be well below 10\% of the solar radius, both by periodic geometric surveys, as well as, twice per year, by filming the Sun with an optical telescope and camera attached to the magnet, and whose optical axis has been set parallel to it. The effect of refraction in the atmosphere, relevant for photons, but not for axions, is properly taken into account. At both ends of the magnet, different x-ray detectors have been searching for photons coming from axion conversion inside the magnet when it is pointing to the Sun. During non-tracking time, calibrations are performed and detectors record background data.

The data presented here were taken with three detection systems. On the sunset (SS) side of the magnet, two gas-based low-background detectors (SS1 and SS2) read by Micromegas planes~\cite{Giomataris:1995fq}, were directly connected to each of the magnet pipes. On the sunrise (SR) side of the magnet, an improved Micromegas detector was situated at the focal plane of the new XRT.
The detectors were small gaseous time projection chambers of 3~cm drift and were filled with 1.4 bar Argon-2\%isobutane mixture. Their cathodes were 4~$\mu$m-thick mylar windows that face the magnet pipe vacuum and hold the pressure difference while being transparent to x-rays. The detector parts were built with carefully selected low-radioactivity materials, and surrounded by passive (copper and lead) and active (5~cm-thick plastic scintillators) shielding. The Micromegas readouts were built with the microbulk technique~\cite{Andriamonje:2010zz}, out of copper and kapton, and were patterned with 500~$\mu$m pixels interconnected in $x$ and $y$ directions~\cite{Aune:2013pna}. These design choices are the outcome of a long-standing effort to understand and reduce background sources in these detectors \cite{Irastorza:2015geo,Aznar:2015iia}. This effort has led to the best background levels ($\sim$10$^{-6}$~keV$^{-1}$~cm$^{-2}$~s$^{-1}$) ever obtained in CAST.

The XRT installed in the SR system was a 1.5~m focal-length telescope that follows a cone-approximation Wolter I design. It is comprised of thermally formed glass substrates deposited with Pt/C multilayers to enhance \hbox{x-ray} reflectivity in the 0.5--10~keV band. The techniques and infrastructure used in fabricating the CAST XRT were originally developed~\cite{nustar_optics1} to make the two hard \hbox{x-ray} telescopes that are flying on NASA's NuSTAR satellite~\cite{NuSTAR_main}. The optical prescription and multilayer coatings were optimized when considering factors including:  the physical constraints of the CAST experiment; the predicted axion spectrum; and the quantum efficiency of the Micromegas detector~\cite{axion_XRT1}. The point spread function (PSF) and effective area (i.e., throughput) of the XRT were calibrated at the PANTER x-ray test facility at MPE in Munich in July 2016. These calibration data were incorporated into Monte Carlo geometric ray-trace simulations to determine the expected 2D distribution of solar axion-induced photons, that is shown in Fig.~\ref{fig:hitmaps}. Although there is a slight energy dependence on the PSF (the XRT focuses better at higher x-ray energy), more than 50\% of the flux is always concentrated in a few mm$^2$ area, effectively reducing the background to levels down to $\sim$0.003~counts/h. In addition, the combined XRT and detector system was regularly calibrated in CAST using an x-ray source placed $\sim$12~m away from the optics (at the SS side of the magnet). One such calibration is shown in Fig.~\ref{fig:hitmaps}, together with the expected 2D distribution from the ray-trace simulation. These contours are different from the ones expected from axion-induced photons (shown in Fig.~\ref{fig:hitmaps}) due to different angular size and distance of the source. The matching between data and simulations confirms our good understanding of the optics performance. This is the first time an XRT has been designed and built specifically for axion physics and operated together with a Micromegas detector at its focal point~\cite{Aznar:2015iia}. This experience is particularly valuable to develop a next-generation scaled-up helioscope.

\begin{table*}[t] \centering \footnotesize
\begin{tabular}{lccccccc}
\\ \hline\hline
\multicolumn{1}{l}{Data set} & Detector & Year & \multicolumn{1}{c}{Tracking
exposure} & \multicolumn{1}{c}{Background exposure} & \multicolumn{2}{c}{Measured count rates ($\pm 1\sigma$ error)}  \\
%\multicolumn{1}{c}{$(g^4_{a\gamma})_{\rm best fit}$ ($\pm 1\sigma$
%error)} & \multicolumn{1}{c}{$\chi^2_{\rm null}$/d.o.f} &
%\multicolumn{1}{c}{$\chi^2_{\rm min}$/d.o.f} &
%\multicolumn{1}{c}{$g_{a\gamma}$(95\%)} \\
& & & (h) & (h) & \multicolumn{2}{c}{($10^{-6}$~\IGI{keV$^{-1}$cm$^{-2}$s$^{-1}$)}}\\
& & & & & Tracking & Background \\ \hline
\\ A & SS1 & 2013 & 92.5 & 1700.0 & $0.79\pm0.18$ & $0.81\pm0.04$ \\
B & SS2 & 2013 & 86.5 & 1407.8 & $1.37\pm0.24$ & $1.48\pm0.06$ \\
C & SS1 & 2014 & 118.0 & 1854.0 & $0.94\pm0.17$ & $1.03\pm0.05$ \\
D & SS2 & 2014 & 118.1 & 1819.6 & $0.97\pm0.18$ & $1.05\pm0.05$ \\
E & SS1 & 2015 & 79.5 & 1237.6 & $0.77\pm0.18$ & $0.89\pm0.05$ \\
F & SS1 & 2015 & 49.7 & 783.1 & $1.77\pm0.36$ & $1.65\pm0.09$ \\
G & SS1 & 2015 & 83.5 & 1431.5 & $1.32\pm0.25$ & $1.10\pm0.05$ \\
H & SS2 & 2015 & 81.3 & 1236.2 & $0.70\pm0.18$ & $0.89\pm0.05$ \\
I & SS2 & 2015 & 51.3 & 800.2 & $1.04\pm0.27$ & $1.59\pm0.08$ \\
J & SS2 & 2015 & 82.0 & 1409.2 & $0.91\pm0.20$ & $0.90\pm0.05$ \\ \hline
K & SR & 2014 & 69.8 & 1379.4 & 0 counts & $0.25 \pm 0.05$ counts \\
L & SR & 2015 & 220.4 & 4125.4 & 3 counts & $ 0.77 \pm 0.15 $ counts \\
%B & SS2 & 2013 &  43.8 & 431.4 & $-1.4\pm 4.5$ & 12.5/14 & 12.4/13 &
%$1.67$\\
%C & 11.5 & 121.0 & $2.5\pm 8.8$ & 6.2/14 & 6.1/13
%& $2.09$
%\\ MM set C & 21.8 & 251.0 & $-9.4\pm 6.5$
%& 12.8/14 & 10.7/13 & $1.67$
%\\ CCD &
%121.3 & 1233.5 & $0.4 \pm 1.0$ & 28.6/20 & 28.5/19 & $1.23$\\
%\hline
\multicolumn{3}{r}{Total tracking exposure (h):}  & \textbf{1132.6} &  &  &  \\ \hline
\end{tabular}
\caption{Tracking and background exposure, as well as the integrated 2--7 keV measured count rate, for both tracking and background data, for each of the data sets included in our result. Note that for rows K and L background levels are expressed in units of total counts in the (95\% signal-enclosing) spot area during the corresponding tracking exposure.\label{tab:datasets}}
\end{table*}

\begin{figure*}[t]
\centering
\includegraphics[width=1.62\columnwidth]{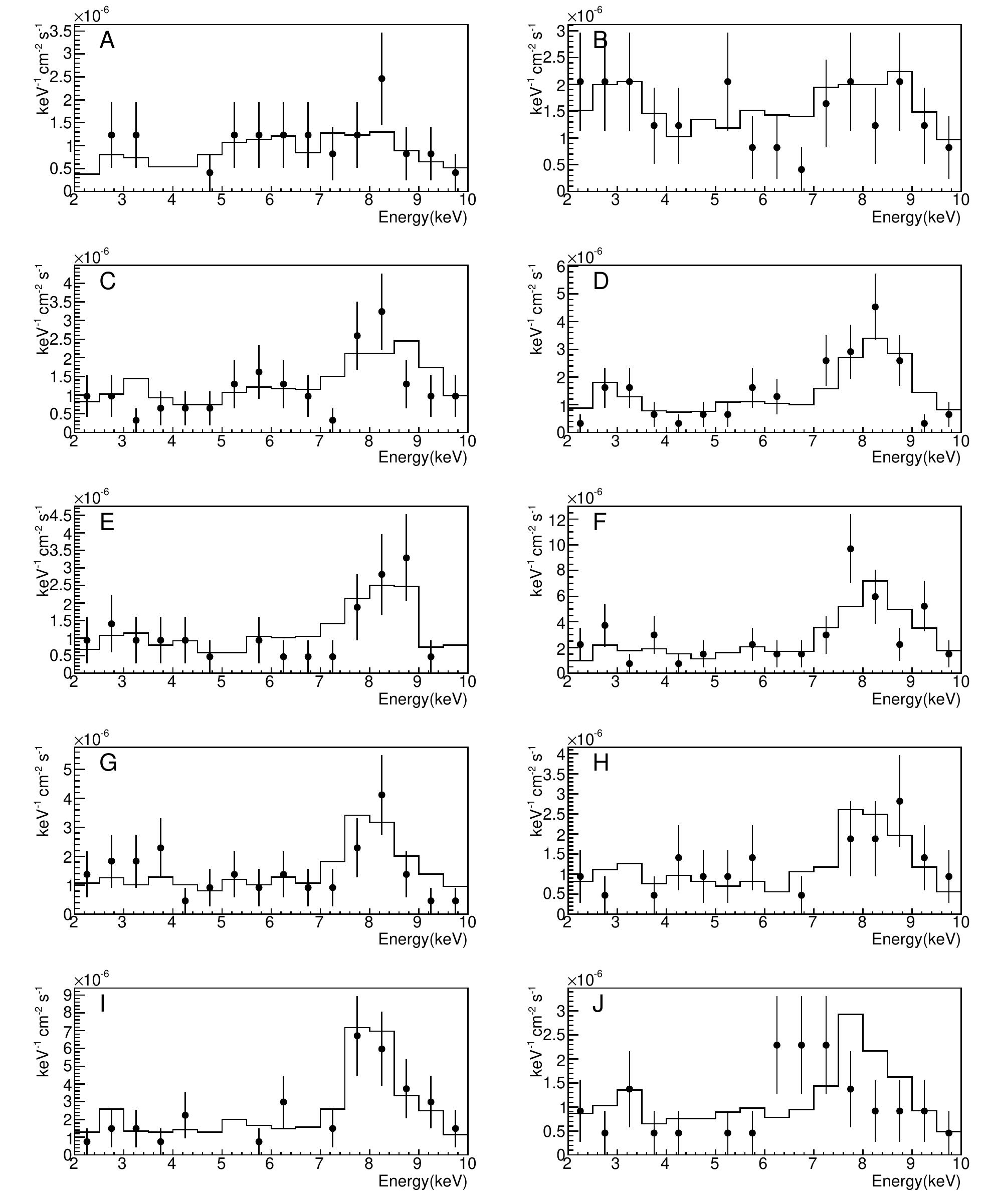}
\caption{Spectra of tracking (dots) and background (solid line) data for each of the sunset datasets. \IGI{The error bars correspond to the 1-$\sigma$ statistical fluctuation of each bin content following Poissonian statistics}. The error bars of the background data are omitted for the sake of clarity, although they are typically $\sim$3 times smaller than the tracking data error bars.}
\label{fig:spectra}
\end{figure*}

Our results correspond to $1132.6$ hour$\times$detector of data taken in axion-sensitive conditions (i.e.\ magnet powered and pointing to the Sun) with the aforementioned detectors in 2013, 2014 and 2015. In 2013, only SS detectors were operative, while in 2014 and 2015 both the SS detectors and the new SR system, installed in CAST in September 2014, took data. The data are divided in sets as shown in table~\ref{tab:datasets}, according to detector and the year of the data-taking campaign.The 2015 SS detectors data are further divided in three sets (E, F, G and H, I, J for SS1 and SS2 detectors respectively) due to an accidental variation in the detector configuration: one of the muon veto remained inoperative for about one month, leading to a different background rate during that period. Background levels are defined independently for every data set using data acquired during non-tracking periods. These data have typically $\sim$10 times more exposure than tracking data and consequently background levels have $\sim$3 times smaller statistical error bars. Data shown in the tables and figures always refer to levels after processing. Raw data from the Micromegas detectors undergo an offline filtering process, detailed elsewhere~\cite{Aune:2013pna}, based on topological information of the event (e.g.\ number of ionization clusters recorded in the chamber, or longitudinal and transversal spread of the signal), to keep only signal-like (i.e.\ x-ray-like) events. The effect of this filtering on raw background levels at low energies is about a factor $\sim$100, while the signal efficiency stays at \hbox{60--70\%}, depending on the event energy, as determined experimentally by careful calibrations with an x-ray tube at different representative energies~\cite{GraciaGarza:2015sos}. In addition, an anti-coincidence condition with the external plastic muon vetoes is applied, leading to an additional factor $\sim$2 rejection (see Fig.~\ref{fig:hitmaps}).

\begin{figure}[b]
\centering
\includegraphics[width=0.8\columnwidth]{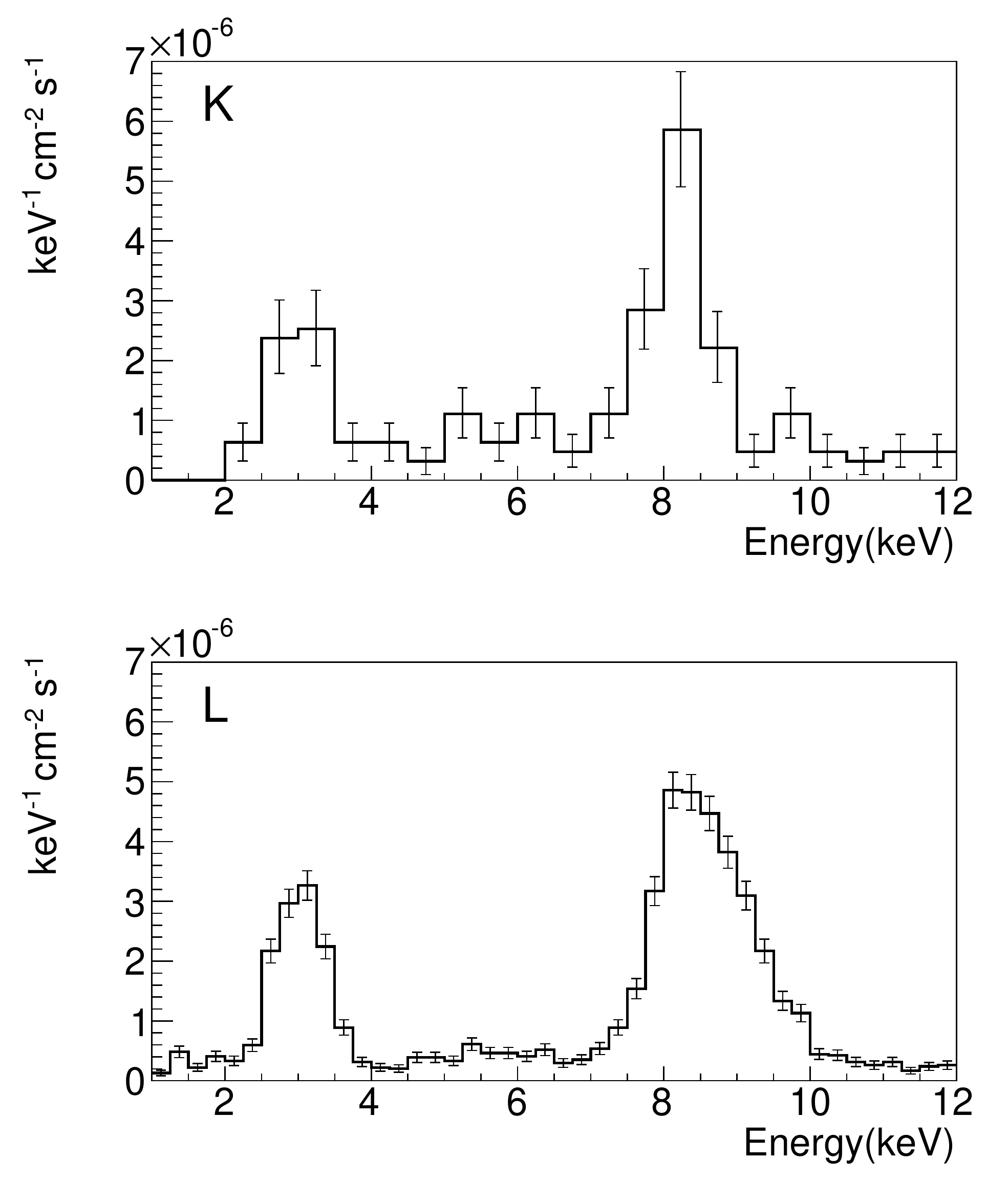}
\caption{Energy spectrum of background data in the sunrise detector (both K and L datasets). \IGI{The error bars correspond to the 1-$\sigma$ statistical fluctuation of each bin content following Poissonian statistics}.}
\label{fig:srback}
\end{figure}

\IGI{The energy range of interest (RoI) is set between 2 and 7 keV, the band that contains most of the expected signal. The low-energy bound is safely above the effective energy threshold of the detectors (which are around $\sim$1 and $\sim$1.5 keV, respectively for SR and SS detectors), and the high-energy bound prevents contamination from the prominent $\sim$8 keV Cu fluorescence peak observed in the background. The measured tracking and background levels, integrated in this RoI, are presented in table~\ref{tab:datasets} for each of the data sets}. Figure~\ref{fig:spectra} shows the spectral distribution of all SS data sets. The background spectra of the two SR data sets are shown in Fig.~\ref{fig:srback}. In these latter plots, data from all the detector area---also outside the signal spot---are included to increase the statistics of the spectra, however only data from the spot area are used to define the background in the analysis. These measured background levels are primarily attributed to cosmic muon-induced secondaries that are not properly tagged (the muon veto coverage of the shielding is not complete due to spatial constraints of the CAST setups) as well as a remaining environmental $\gamma$-induced background population that reaches the detector despite the shielding, probably associated with the small solid angle facing the magnet that is not possible to shield. This insight is corroborated by the $\sim$3~keV Ar and $\sim$8 keV Cu fluorescence peaks observed in the background spectra (better seen in the SR plot of Fig.~\ref{fig:srback}, due to the better energy resolution of this detector), and will be the basis of future improvements of the detector~\cite{Irastorza:2015geo}.

Figure~\ref{fig:hitmaps} shows the 2D distribution of detected events, both in background and tracking data, in the SR detector, superimposed on the region where the signal is expected. The background level in the signal area could be estimated using the data outside the spot or the data in the spot in non-tracking periods. This second method is preferable, as the background level shows a slight increase at the center of the detector (attributed to the detector and shielding geometry). When normalized to the 290~h of tracking data available (data sets K and L in table ~\ref{tab:datasets}) only $1.02 \pm 0.22$ ($2.13 \pm 0.47$) background counts are expected in the 95\% (99\%) signal-enclosing focal spot region\IGI{, where errors indicate 1-$\sigma$ intervals}. The tracking data reveal 3(4) observed counts inside such regions. Their measured energies are 3.05, 2.86, 2.94 and 2.56~keV.

\section{Data analysis and results}

The data analysis follows similar previous analyses of CAST data~\cite{Arik:2011rx,Arik:2013nya,Arik:2015cjv}.
We define an unbinned likelihood function
\begin{equation}
\log{\mathcal{L}} \propto -R_T + \sum_{i}^n \log{ R(E_i,d_i,\textbf{x}_i)}\,,
\end{equation}
where $R_T$ is the expected number of counts from the axion-to-photon conversion in all data sets, integrated over the tracking exposure time and energy of interest. The sum is over each of the $n$ detected counts \IGI{in the energy RoI} during the tracking time, for an expected rate $R(E_i,d_i,\textbf{x}_i)$ as a function of the energy $E_i$, data set $d_i$, and detector coordinates $\textbf{x}_i$ of the event $i$, and given by the expression
\begin{equation}
\IGI{R(E,d,\textbf{x})} = B(E,d) + S(E,d,\textbf{x})\,,
\end{equation}
where $B(E,d)$ is the background level for data set $d$, considered constant in time and $\textbf{x}_i$ within the data set. $S(E,d,\textbf{x})$ is the expected rate from axion conversion in the detector of data set $d$ given by
\begin{equation}
S(E,d,\textbf{x}) = \frac{d \Phi_{a}} {d E} P_{a\rightarrow\gamma} \epsilon (d,E,\textbf{x})\,.
\end{equation}
Here, $\epsilon (d,E,\textbf{x})$ is the detector response for data set $d$, and includes both the $E$-dependent detector efficiency (both hardware and software), and for the SR system also the $E$-dependent optics throughput and the expected signal distribution over $\textbf{x}$ due to the optics PSF shown in Fig.~\ref{fig:hitmaps}. For the SS detector there is no such dependency and  $\epsilon (d,E,\textbf{x})=\epsilon(d,E)$.

Finally, $d \Phi_{a}/dE$ is the differential solar axion flux, which can be parameterized by the expression~\cite{Andriamonje:2007ew}
\begin{equation}
\frac {d \Phi_{a}} {d E} = 6.02 \times 10^{10} g_{10}^{2} \frac{E^{2.481}} {e^{E/1.205}}
\left[ \hbox{cm}^{-2}~\hbox{s}^{-1}~\hbox{keV}^{-1} \right]
\end{equation}
with $g_{10}=g_{a\gamma}/(10^{-10}~\mbox{GeV}^{-1})$ and energy $E$ in keV. The
axion-to-photon conversion probability $P_{a\rightarrow\gamma}$
was given in Eq.~(\ref{eq:ConvProb}).

By numerically maximizing $\log{\mathcal{L}}$ a best-fit value $g_{10,\rm min}^4$ is
obtained. This value is compatible with the absence of a signal in the entire axion mass range
and thus an upper limit on $g_{a\gamma}$ is extracted, by
integrating the Bayesian posterior probability density function (PDF) from zero up to 95\%
of the total PDF area, using a flat prior in $g_{10}^{4}$ for positive values, and zero for the unphysical negative ones. The computed upper limit for
several values of $m_a$ is displayed in red in Fig. \ref{fig:limits}.

\begin{figure*}[t]
\includegraphics[width=1.8\columnwidth]{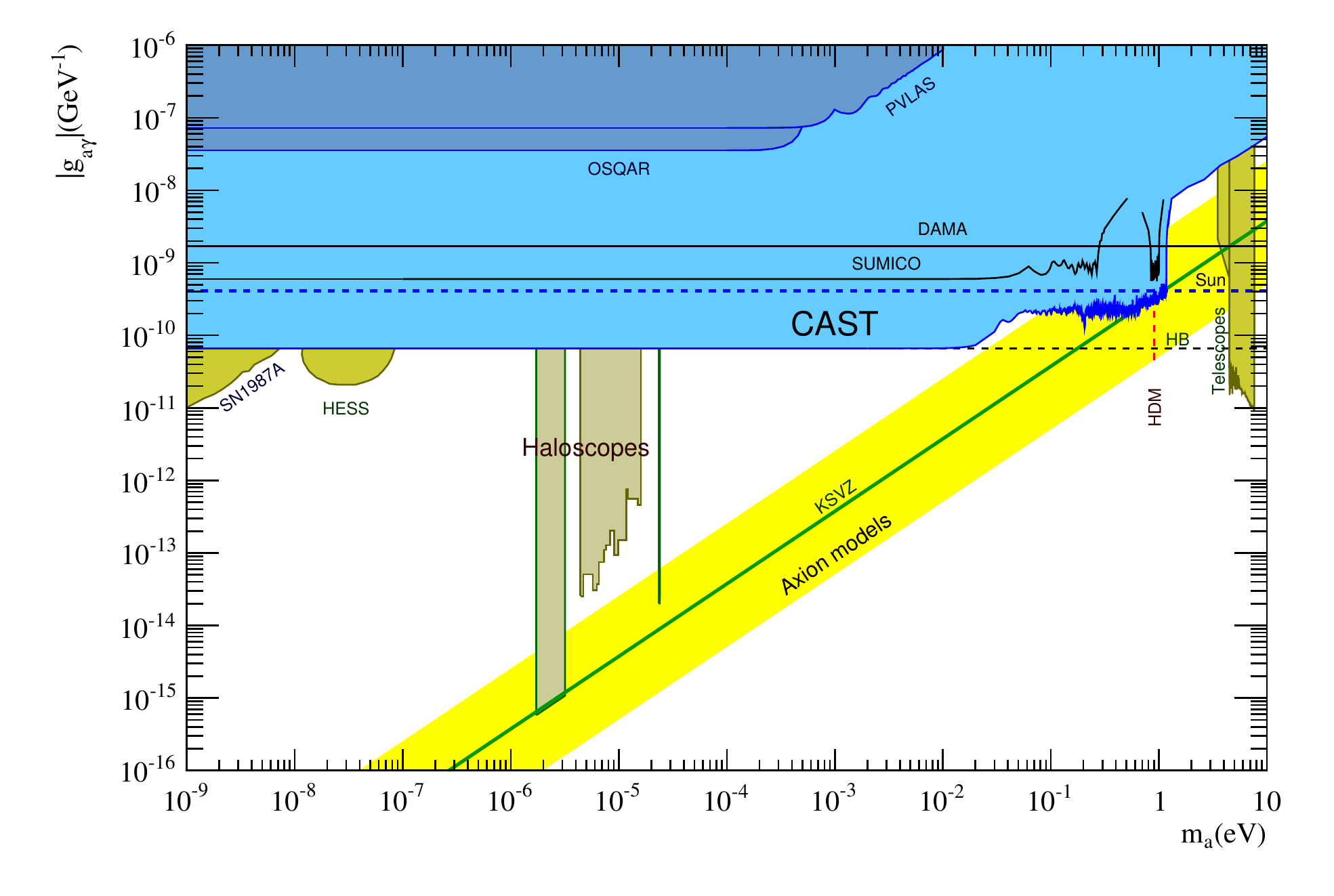}
\caption{Constraints on the two-photon coupling $g_{a\gamma}$ of axions and similar particles depending on
their mass $m_a$. Apart from the CAST limits updated with the result presented here,
we show the results from the previous helioscope Sumico and the best limit from the
Bragg technique (DAMA). Moreover, we show
the latest limits from the laser propagation experiments OSQAR and PVLAS, high-energy photon propagation
in astrophysical $B$-fields (H.E.S.S.), the SN1987A observation, and telescope limits for cosmic axion decay lines. Horizontal dashed lines provide limits from properties of the Sun
and the energy loss of horizontal branch (HB) stars. The vertical dashed line denotes
the cosmic hot dark matter (HDM) limit which only applies to QCD axions.
The haloscope limits assume that axions are the galactic dark matter.
The yellow band of QCD axion models and the green KSVZ line are as in Fig.~\ref{fig:limits}.}
\label{fig:panoramaplot}
\end{figure*}

For mass values below $m_a\alt0.02~{\rm eV}$, the mass independent best fit value is $g_{10,\rm{min}}^{4}=(-0.06^{+0.10}_{-0.07})$, where the errors indicate 1-$\sigma$ intervals. The upper limit in this mass range constitutes our main result:
\begin{equation}\label{eq:final-limit}
g_{a\gamma}<0.66\times10^{-10}~{\rm GeV}^{-1}
\quad\hbox{at 95\% C.L.}
\end{equation}
This constraint considers only statistical fluctuations in the tracking data. Other potentially important
systematic effects would include uncertainties in input parameters like the magnet length and strength, the background levels, the tracking accuracy, the optical alignment of the SR system, and the theoretical uncertainty of the expected signal. All these contributions have been estimated to be negligible in the final result, which is dominated by the statistical error of the low-counting observation. Overall systematic effects are well below 10\% of the quoted result. It is worth noting that the observation of the SR counts represents a statistical fluctuation that slightly worsens the final result with respect to the expected sensitivity (defined as the average exclusion of an ensemble of possible statistical outcomes under the background-only hypothesis), estimated as $g_{10,\rm{average}} \lesssim 0.64$.

\section{Discussion}

The final solar axion run of CAST has provided the
new constraint of Eq.~(\ref{eq:final-limit}) on the axion-photon
coupling strength for $m_a\lesssim 0.02~{\rm eV}$. It is shown
in the wider \hbox{$m_a$--$g_{a\gamma}$} landscape in Fig.~\ref{fig:panoramaplot}.
In particular, axions and photons can also interconvert in astrophysical $B$ fields
that tend to be much weaker than in CAST, but extend over much
larger distances. For example, the non-observation of $\gamma$
rays contemporaneous with the SN~1987A neutrino signal provides a
limit $g_{10}\alt0.053$ for \IGI{$m_a\alt0.44~{\rm neV}$} \cite{Payez:2014xsa}.
This conversion effect could be important also for other astrophysical
and cosmological sources,
especially for the propagation of TeV $\gamma$ rays
and as an explanation for the soft x-ray excess in galaxy clusters
\GGR{(for a recent review see Ref.~\cite{Meyer:2016wrm})}.
In the very low mass domain, propagation experiments for laser beams
in magnetic fields have also explored the
$g_{a\gamma}$--$m_a$ parameter space~\cite{Ballou:2015cka,DellaValle:2015xxa}.

Helioscopes using the Bragg technique can reach larger
axion masses than CAST because the electric fields
in the crystals are inhomogeneous on much smaller scales. Yet because
of their limited size, even the best case shown in Fig.~\ref{fig:panoramaplot}
from DAMA \cite{Bernabei:2001ny} is not competitive.
In the future, CUORE may reach values of
$g_{a\gamma}$ comparable to the CAST $^3$He limits near $m_a\sim 1$~eV, but
extended to larger masses~\cite{Li:2015tsa}.

However, QCD axions in this parameter range would thermalize in the early universe
and provide a hot dark matter fraction. Cosmology provides a constraint corresponding to
$m_a<0.86$~eV (vertical dashed red line in Fig.~\ref{fig:panoramaplot})
which in future may reach $m_a<0.15$~eV \cite{Archidiacono:2013cha}. Therefore, exploring
beyond the 1~eV mass range is only relevant for those axion-like particles which, unlike
QCD axions, do not thermalize efficiently.

Solar axion searches usually
assume that the axion flux is only a small perturbation of the Sun. Actually for $g_{10}\agt 20$ axion losses are so large that one cannot construct self-consistent
solar models \cite{Schlattl:1998fz}. Moreover, the measured solar neutrino flux and helioseismology
require $g_{10}<4.1$ at $3\sigma$ confidence \cite{Vinyoles:2015aba} (dashed blue line in
Fig.~\ref{fig:panoramaplot}), implying that CAST is the only solar
axion search that has gone beyond this recent limit.

A sensitivity comparable to the new CAST limit derives from traditional
stellar energy loss arguments.
In particular, Primakoff losses accelerate the helium-burning phase of horizontal
branch (HB) stars, reducing their number count relative to low-mass
red giants, $R=N_{\rm HB}/N_{\rm RGB}$, in globular clusters \IGI{(see dashed line labelled ``HB'' in Fig.~\ref{fig:panoramaplot})}. The most
recent analysis finds $g_{10}<0.66$ (95\% C.L.) and actually a mild
preference for $g_{10}\sim0.4$, although a detailed budget of
systematic uncertainties is not currently available~\cite{Ayala:2014pea}.

Solar axion searches beyond CAST and at the same time beyond the HB star limit
as a benchmark require
a new effort on a much larger scale, for example the proposed helioscope IAXO
\cite{Armengaud:2014gea}.
For small masses,
new regions in the $g_{a\gamma}$--$m_a$ will be explored with the upcoming
ALPS-II laser propagation experiment at DESY \cite{Bahre:2013ywa}, by higher-statistics
TeV $\gamma$-ray observations, or the $\gamma$-ray signal from a future galactic
supernova \cite{Meyer:2016wrm}. Beyond the $g_{a\gamma}$--$m_a$ parameter space, many attractive
detection opportunities are pursued world wide, notably under the
assumption that axions are the cosmic dark matter \cite{AxDM-Searches}.

After finishing its solar axion program in 2015, CAST itself has turned to
a broad physics program at the low-energy frontier.
In particular, this includes KWISP, a sensitive force detector, and an InGrid detector, both in search of solar chameleons, as well as \hbox{CAST-CAPP} and RADES, implementing long aspect-ratio microwave cavities in the CAST magnet to search for dark-matter axions around $m_a\sim20$~$\mu$eV~\cite{spsc_report}.

Whichever of the many new developments world wide will lead to the detection of axions
or similar particles, in the \hbox{$g_{a\gamma}$--$m_a$}
parameter space our new CAST result will remain the benchmark for many years
to come and guide future explorations.

\section{Acknowledgments}

We thank CERN for hosting the experiment and for
technical support to operate the magnet and cryogenics. \IGI{We thank V. Burwitz, G. Hartner of the MPE PANTER x-ray test facility for providing the opportunity to calibrate the x-ray telescope and for assistance in collecting and analyzing the characterization data}. We
acknowledge support from NSERC (Canada), MSE (Croatia) and Croatian
Science Foundation under the project IP-2014-09-3720,
CEA (France), BMBF (Germany) under the grant numbers 05 CC2EEA/9 and 05 CC1RD1/0 and DFG (Germany) under
grant numbers HO 1400/7-1 and EXC-153, GSRT (Greece), NSRF:
Heracleitus II, RFFR (Russia), the Spanish Ministry of Economy and
Competitiveness (MINECO) under Grants No.\ FPA2011-24058 and
No.\ FPA2013-41085-P (grants partially funded by the European
Regional Development Fund, ERDF/FEDER), the European Research Council
(ERC) under grant ERC-2009-StG-240054 (T-REX), Turkish Atomic Energy
Authority (TAEK), NSF (USA) under Award No.\ 0239812, NASA under the grant number NAG5-10842, and IBS (Korea) with code IBS-R017-D1-2017-a00. Part of this
work was performed under the auspices of the U.S.\ Department of
Energy by Lawrence Livermore National Laboratory under Contract
No.\ DE-AC52-07NA27344.

\IGI{
\section*{Data Availability Statement}
The data that support the plots within this paper and other findings of this study are available from the corresponding author upon reasonable request.
}

\IGI{
\section*{Contributions}
S.A., J.F.C., T.Da., I.G., E.F.R., F.J.I., J.A.G., I.G.I., J.G.G. and T.P. conceived, designed and built the Micromegas x-ray detectors. F.C., T.De., C.J.H., A.J., M.J.P., J.R. and J.K.V. conceived, designed, calibrated and built the x-ray telescope. J.F.C., T.Da., F.J.I., J.A.G., J.G.G., E.R.C. and J.R. installed, calibrated and operated optics and detectors in the experiment. T.Da., F.J.I., J.A.G., J.G.G., I.G.I. and G.L. conceived and implemented the low-background measures on the detection systems. K.B., M.D., J.M.L., L.S., T.V., S.C.Y. and K.Z. operated and maintained the magnet, vacuum, slow control and other ancillary systems, as well as the interfaces with the detection systems. M.D., T.V. and M.K. organized the periodical pointing calibration of the tracking system. B.L., T.V., M.D. and K.Z. organized the data taking runs and the general operation of the experiment. M.J.P. performed the optics ray-trace models. J.A.G., F.J.I and I.G.I. monitored and analyzed the detector data. I.G.I. coordinated the high level analysis and made the plots. I.G.I., B.L., M.J.P. and G.R. wrote the manuscript.
All authors contributed to the data taking shifts and the operation and maintenance of the experiment at a whole, and all authors commented on the manuscript.
}


\begin{thebibliography}{99}

\bibitem{Olive:2016xmw}
  C.~Patrignani {\it et al.} (Particle Data Group),
  Review of particle physics,
  Chin.\ Phys.\ C {\bf 40}, 100001 (2016).
  %doi:10.1088/1674-1137/40/10/100001
  %%CITATION = doi:10.1088/1674-1137/40/10/100001;%%
  %207 citations counted in INSPIRE as of 27 Dec 2016

%\cite{Jaeckel:2010ni}
\bibitem{Jaeckel:2010ni}
  J.~Jaeckel and A.~Ringwald,
  The low-energy frontier of particle physics,
  Ann.\ Rev.\ Nucl.\ Part.\ Sci.\  {\bf 60}, 405 (2010).
  %[arXiv:1002.0329 [hep-ph]].
  %%CITATION = ARXIV:1002.0329;%%

\bibitem{AxionBook}
  M.~Kuster, G.~Raffelt and B.~Beltr\'an (eds.),
  Axions: Theory, Cosmology, and Experimental Searches,
  Lect.\ Notes Phys.\ {\bf 741}, 1--237 (2008).

\bibitem{AxDM-Searches}
  P.~W.~Graham, I.~G.~Irastorza, S.~K.~Lamoreaux, A.~Lindner and K.~A.~van Bibber,
  Experimental searches for the axion and axion-like particles,
  Ann.\ Rev.\ Nucl.\ Part.\ Sci.\  {\bf 65}, 485 (2015).
  %doi:10.1146/annurev-nucl-102014-022120
  %[arXiv:1602.00039].
  %%CITATION = doi:10.1146/annurev-nucl-102014-022120;%%
  %22 citations counted in INSPIRE as of 24 Nov 2016

\bibitem{Patras}
  For recent experimental and theoretical developments see
  the presentations at the annual ``Patras'' workshop series
  https://axion-wimp2017.desy.de/e28952/

\bibitem{Forces}
  A.~Arvanitaki and A.~A.~Geraci,
  Resonantly detecting axion-mediated forces with nuclear magnetic resonance,
  Phys.\ Rev.\ Lett.\  {\bf 113}, 161801 (2014).
  %doi:10.1103/PhysRevLett.113.161801
  %[arXiv:1403.1290 [hep-ph]].
  %%CITATION = doi:10.1103/PhysRevLett.113.161801;%%
  %38 citations counted in INSPIRE as of 20 Jan 2017

\bibitem{Superradiance}
  R.~Brito, V.~Cardoso and P.~Pani,
  {\em Superradiance: Energy Extraction, Black-Hole Bombs and Implications for Astrophysics and Particle
  Physics},
  Lect.\ Notes Phys.\  {\bf 906}, 1 (2015).
  %doi:10.1007/978-3-319-19000-6
  %[arXiv:1501.06570 [gr-qc]].
  %%CITATION = doi:10.1007/978-3-319-19000-6;%%
  %73 citations counted in INSPIRE as of 20 Jan 2017

%\cite{Ayala:2014pea}
\bibitem{Ayala:2014pea}
  A.~Ayala, I.~Dom{\'\i}nguez, M.~Giannotti, A.~Mirizzi and O.~Straniero,
  Revisiting the bound on axion-photon coupling from globular clusters,
  Phys.\ Rev.\ Lett.\ {\bf 113}, 191302 (2014).
  %[arXiv:1406.6053 [astro-ph.SR]].
  %%CITATION = ARXIV:1406.6053;%%
  %7 citations counted in INSPIRE as of 10 Dec 2014

%\cite{Giannotti:2015kwo}
\bibitem{Giannotti:2015kwo}
  M.~Giannotti, I.~Irastorza, J.~Redondo and A.~Ringwald,
  Cool WISPs for stellar cooling excesses,
  JCAP {\bf 1605}, 057 (2016).
  %doi:10.1088/1475-7516/2016/05/057
  %[arXiv:1512.08108 [astro-ph.HE]].
  %%CITATION = doi:10.1088/1475-7516/2016/05/057;%%
  %5 citations counted in INSPIRE as of 10 Oct 2016

%\cite{Sikivie:1983ip}
\bibitem{Sikivie:1983ip}
  P.~Sikivie,
  Experimental tests of the invisible axion,
  Phys.\ Rev.\ Lett.\  {\bf 51}, 1415 (1983);
  Erratum {\it ibid.}\  {\bf 52}, 695 (1984).
  %%CITATION = PRLTA,51,1415;%%
  %677 citations counted in INSPIRE as of 12 May 2014

%\cite{Raffelt:1987im}
\bibitem{Raffelt:1987im}
  G.~Raffelt and L.~Stodolsky,
  Mixing of the photon with low mass particles,
  Phys.\ Rev.\ D {\bf 37}, 1237 (1988).
  %doi:10.1103/PhysRevD.37.1237
  %%CITATION = doi:10.1103/PhysRevD.37.1237;%%
  %407 citations counted in INSPIRE as of 09 Oct 2016

%\cite{Zioutas:2004hi}
\bibitem{Zioutas:2004hi}
  K.~Zioutas {\it et al.} (CAST Collaboration),
  First results from the CERN Axion Solar Telescope (CAST),
  Phys.\ Rev.\ Lett.\  {\bf 94}, 121301 (2005).
  %[arXiv:hep-ex/0411033].
  %%CITATION = PRLTA,94,121301;%%

%\cite{Andriamonje:2007ew}
\bibitem{Andriamonje:2007ew}
  S.~Andriamonje {\it et al.} (CAST Collaboration),
  An improved limit on the axion-photon coupling from the CAST experiment,
  JCAP {\bf 0704}, 010 (2007).
  %[arXiv:hep-ex/0702006].
  %%CITATION = JCAPA,0704,010;%%

%\cite{Arik:2008mq}
\bibitem{Arik:2008mq}
  E.~Arik {\it et al.} (CAST Collaboration),
  Probing eV-scale axions with CAST,
  JCAP {\bf 0902}, 008 (2009).
  %[arXiv:0810.4482 [hep-ex]].
  %%CITATION = JCAPA,0902,008;%%

%\cite{Arik:2015cjv}
\bibitem{Arik:2015cjv}
  M.~Arik {\it et al.} (CAST Collaboration),
  New solar axion search using the CERN Axion Solar Telescope with $^4$He filling,
  Phys.\ Rev.\ D {\bf 92}, 021101 (2015).
  %doi:10.1103/PhysRevD.92.021101
  %[arXiv:1503.00610 [hep-ex]].
  %%CITATION = doi:10.1103/PhysRevD.92.021101;%%
  %12 citations counted in INSPIRE as of 08 Oct 2016

%\cite{Arik:2011rx}
\bibitem{Arik:2011rx}
  M.~Arik {\it et al.} (CAST Collaboration),
  CAST search for sub-eV mass solar axions with $^3$He buffer gas,
  Phys.\ Rev.\ Lett.\  {\bf 107}, 261302 (2011).
  %[arXiv:1106.3919 [hep-ex]].
  %%CITATION = ARXIV:1106.3919;%%

%\cite{Arik:2013nya}
\bibitem{Arik:2013nya}
  M.~Arik {\it et al.} (CAST Collaboration),
  CAST solar axion search with $^3$He buffer gas: Closing the hot dark matter gap,
  Phys.\ Rev.\ Lett.\  {\bf 112}, 091302 (2014).
  %[arXiv:1307.1985 [hep-ex]].
  %%CITATION = ARXIV:1307.1985;%%
  %12 citations counted in INSPIRE as of 11 May 2014

\bibitem{SUMICO}
  Y.~Inoue,
  Y.~Akimoto, R.~Ohta, T.~Mizumoto, A.~Yamamoto and M.~Minowa,
  Search for solar axions with mass around 1~eV using coherent conversion of
  axions into photons,
  Phys.\ Lett.\  B {\bf 668}, 93 (2008).
  %[arXiv:0806.2230 [astro-ph]].
  %%CITATION = PHLTA,B668,93;%%

\bibitem{Paschos:1993yf}
  E.~A.~Paschos and K.~Zioutas,
  A proposal for solar axion detection via Bragg scattering,
  Phys.\ Lett.\ B {\bf 323}, 367 (1994).
  %doi:10.1016/0370-2693(94)91233-5
  %%CITATION = doi:10.1016/0370-2693(94)91233-5;%%
  %75 citations counted in INSPIRE as of 04 Jan 2017

%\cite{Bernabei:2001ny}
\bibitem{Bernabei:2001ny}
  R.~Bernabei {\it et al.},
  Search for solar axions by Primakoff effect in NaI crystals,
  Phys.\ Lett.\ B {\bf 515}, 6 (2001).
  %doi:10.1016/S0370-2693(01)00840-1
  %%CITATION = doi:10.1016/S0370-2693(01)00840-1;%%
  %133 citations counted in INSPIRE as of 09 Oct 2016

%\cite{Andriamonje:2009dx}
\bibitem{Andriamonje:2009dx}
  S.~Andriamonje {\it et al.} (CAST Collaboration),
  Search for 14.4-keV solar axions emitted in the M1-transition of Fe-57 nuclei with CAST,
  JCAP {\bf 0912}, 002 (2009).
  %doi:10.1088/1475-7516/2009/12/002
  %[arXiv:0906.4488 [hep-ex]].
  %%CITATION = doi:10.1088/1475-7516/2009/12/002;%%
  %32 citations counted in INSPIRE as of 09 Oct 2016

%\cite{Andriamonje:2009ar}
\bibitem{Andriamonje:2009ar}
  S.~Andriamonje {\it et al.} (CAST Collaboration),
  Search for solar axion emission from $^7$Li and D$(p,\gamma){}^3$He nuclear decays with
  the CAST $\gamma$-ray calorimeter,
  JCAP {\bf 1003}, 032 (2010).
  %doi:10.1088/1475-7516/2010/03/032
  %[arXiv:0904.2103 [hep-ex]].
  %%CITATION = doi:10.1088/1475-7516/2010/03/032;%%
  %26 citations counted in INSPIRE as of 09 Oct 2016

%\cite{Barth:2013sma}
\bibitem{Barth:2013sma}
  K.~Barth {\it et al.} (CAST Collaboration),
  CAST constraints on the axion-electron coupling,
  JCAP {\bf 1305}, 010 (2013).
  %doi:10.1088/1475-7516/2013/05/010
  %[arXiv:1302.6283 [astro-ph.SR]].
  %%CITATION = doi:10.1088/1475-7516/2013/05/010;%%
  %27 citations counted in INSPIRE as of 09 Oct 2016

%\cite{Anastassopoulos:2015yda}
\bibitem{Anastassopoulos:2015yda}
  V.~Anastassopoulos {\it et al.} (CAST Collaboration),
  Search for chameleons with CAST,
  Phys.\ Lett.\ B {\bf 749}, 172 (2015).
  %doi:10.1016/j.physletb.2015.07.049
  %[arXiv:1503.04561 [astro-ph.SR]].
  %%CITATION = doi:10.1016/j.physletb.2015.07.049;%%
  %5 citations counted in INSPIRE as of 09 Oct 2016

%\cite{Redondo:2015iea}
\bibitem{Redondo:2015iea}
  J.~Redondo,
  Atlas of solar hidden photon emission,
  JCAP {\bf 1507}, 024 (2015).
  %doi:10.1088/1475-7516/2015/07/024
  %[arXiv:1501.07292 [hep-ph]].
  %%CITATION = doi:10.1088/1475-7516/2015/07/024;%%
  %6 citations counted in INSPIRE as of 09 Oct 2016

%\cite{DiLuzio:2016sbl}
\bibitem{DiLuzio:2016sbl}
  L.~Di Luzio, F.~Mescia and E.~Nardi,
  Redefining the axion window,
  Phys.\ Rev.\ Lett.\  {\bf 118}, 031801 (2017).
  %doi:10.1103/PhysRevLett.118.031801
  %[arXiv:1610.07593 [hep-ph]].
  %%CITATION = doi:10.1103/PhysRevLett.118.031801;%%
  %4 citations counted in INSPIRE as of 11 Mar 2017

%\cite{Autiero:2007uf}
\bibitem{Autiero:2007uf}
  D.~Autiero {\it et al.},
  The CAST Time Projection Chamber,
  New J.\ Phys.\  {\bf 9}, 171 (2007).
  %doi:10.1088/1367-2630/9/6/171
  %[physics/0702189 [physics.ins-det]].
  %%CITATION = doi:10.1088/1367-2630/9/6/171;%%
  %43 citations counted in INSPIRE as of 04 Jan 2017

\bibitem{Abbon:2007ug}
  P.~Abbon {\it et al.},
  The Micromegas detector of the CAST experiment,
  New J.\ Phys.\  {\bf 9}, 170 (2007).
  %doi:10.1088/1367-2630/9/6/170
  %[physics/0702190 [PHYSICS]].
  %%CITATION = doi:10.1088/1367-2630/9/6/170;%%
  %68 citations counted in INSPIRE as of 04 Jan 2017

%\cite{Kuster:2007ue}
\bibitem{Kuster:2007ue}
  M.~Kuster {\it et al.},
  The X-ray Telescope of CAST,
  New J.\ Phys.\  {\bf 9}, 169 (2007)
%  doi:10.1088/1367-2630/9/6/169
 % [physics/0702188 [physics.ins-det]].
  %%CITATION = doi:10.1088/1367-2630/9/6/169;%%
  %58 citations counted in INSPIRE as of 16 Mar 2017

%\cite{Armengaud:2014gea}
\bibitem{Armengaud:2014gea}
  E.~Armengaud {\it et al.},
  Conceptual design of the International Axion Observatory (IAXO),
  JINST {\bf 9}, T05002 (2014).
  %[arXiv:1401.3233 [physics.ins-det]].
  %%CITATION = ARXIV:1401.3233;%%
  %9 citations counted in INSPIRE as of 10 Dec 2014

%\cite{Zioutas:1998cc}
\bibitem{Zioutas:1998cc}
  K.~Zioutas {\it et al.},
  A Decommissioned LHC model magnet as an axion telescope,
  Nucl.\ Instrum.\ Meth.\ A {\bf 425}, 480-489 (1999).
  %doi:10.1016/S0168-9002(98)01442-9
  %[astro-ph/9801176].
  %%CITATION = doi:10.1016/S0168-9002(98)01442-9;%%
  %110 citations counted in INSPIRE as of 04 Jan 2017

\bibitem{Giomataris:1995fq}
  Y.~Giomataris, P.~Rebourgeard, J.~P.~Robert and G.~Charpak,
  MICROMEGAS: A High granularity position sensitive gaseous detector for high particle flux environments,
  Nucl.\ Instrum.\ Meth.\ A {\bf 376}, 29-35 (1996).
  %doi:10.1016/0168-9002(96)00175-1
  %%CITATION = doi:10.1016/0168-9002(96)00175-1;%%
  %508 citations counted in INSPIRE as of 24 Nov 2016

%\bibitem{Aznar:2013jwa}
%  F.~Aznar {\it et al.},
%  Assessment of material radiopurity for Rare Event experiments using Micromegas,
%  JINST {\bf 8}, C11012 (2013).
%  %doi:10.1088/1748-0221/8/11/C11012
%  %%CITATION = doi:10.1088/1748-0221/8/11/C11012;%%
%  %16 citations counted in INSPIRE as of 24 Nov 2016

%\bibitem{Cebrian:2010ta}
 % S.~Cebrian {\it et al.},
  %``Radiopurity of Micromegas readout planes,''
  %Astropart.\ Phys.\  {\bf 34}, 354 (2011).
  %doi:10.1016/j.astropartphys.2010.09.003
  %[arXiv:1005.2022].
  %%CITATION = doi:10.1016/j.astropartphys.2010.09.003;%%
  %58 citations counted in INSPIRE as of 24 Nov 2016

\bibitem{Andriamonje:2010zz}
  S.~Andriamonje {\it et al.},
  Development and performance of Microbulk Micromegas detectors,
  JINST {\bf 5}, P02001 (2010).
  %doi:10.1088/1748-0221/5/02/P02001
  %%CITATION = doi:10.1088/1748-0221/5/02/P02001;%%
  %82 citations counted in INSPIRE as of 24 Nov 2016

\bibitem{Aune:2013pna}
  S.~Aune {\it et al.},
  Low background x-ray detection with Micromegas for axion research,
  JINST {\bf 9}, P01001 (2014).
  %doi:10.1088/1748-0221/9/01/P01001
  %[arXiv:1310.3391].
  %%CITATION = doi:10.1088/1748-0221/9/01/P01001;%%
  %27 citations counted in INSPIRE as of 24 Nov 2016

\bibitem{Irastorza:2015geo}
  I.~G.~Irastorza {\it et al.},
  Gaseous time projection chambers for rare event detection: Results from the T-REX project.
  II. Dark matter,
  JCAP {\bf 1601} (2016) 034;
  Erratum {\em ibid.} {\bf 1605}, E01 (2016).
  %doi:10.1088/1475-7516/2016/05/E01, 10.1088/1475-7516/2016/01/034
  %[arXiv:1512.06294].
  %%CITATION = doi:10.1088/1475-7516/2016/05/E01, 10.1088/1475-7516/2016/01/034;%%
  %5 citations counted in INSPIRE as of 24 Nov 2016

%\bibitem{Aune:2013nza}
 % S.~Aune {\it et al.},
  %``X-ray detection with Micromegas with background levels below 10$^{-6}$ keV$^{-1}$cm$^{-2}$s$^{-1}$,''
  %JINST {\bf 8}, C12042 (2013).
  %doi:10.1088/1748-0221/8/12/C12042
  %[arXiv:1312.4282].
  %%CITATION = doi:10.1088/1748-0221/8/12/C12042;%%
  %17 citations counted in INSPIRE as of 24 Nov 2016

%\cite{Aznar:2015iia}
\bibitem{Aznar:2015iia}
  F.~Aznar {\it et al.},
  A Micromegas-based low-background x-ray detector coupled to a slumped-glass telescope for axion
  research,
  JCAP {\bf 1512}, 008 (2015).
%  doi:10.1088/1475-7516/2015/12/008
 % [arXiv:1509.06190 [physics.ins-det]].
  %%CITATION = doi:10.1088/1475-7516/2015/12/008;%%
  %6 citations counted in INSPIRE as of 13 Jan 2017

\bibitem{nustar_optics1}
  J.~E.~Koglin {\it et al.},
  NuSTAR hard x-ray optics design and performance,
  Proc.\ SPIE Int.\ Soc.\ Opt.\ Eng.\ {\bf 7437}, 74370C (2009).

%\cite{Harrison:2013md}
\bibitem{NuSTAR_main}
  F.~A.~Harrison {\it et al.},
  The Nuclear Spectroscopic Telescope Array (NuSTAR) High-Energy X-Ray Mission,
  Astrophys.\ J.\  {\bf 770}, 103 (2013).
  %doi:10.1088/0004-637X/770/2/103
  %[arXiv:1301.7307].
  %%CITATION = doi:10.1088/0004-637X/770/2/103;%%
  %349 citations counted in INSPIRE as of 10 Dec 2016

\bibitem{axion_XRT1}
  A.~C.~Jakobsen, M.~J.~Pivovaroff and F.~E.~Christensen,
  X-ray optics for axion helioscopes,
  Proc.\ SPIE Int.\ Soc.\ Opt.\ Eng., {\bf 8861}, 886113 (2013).

%\cite{GraciaGarza:2015sos}
\bibitem{GraciaGarza:2015sos}
  J.~Gracia Garza,
  Micromegas for the search of solar axions in CAST and low-mass WIMPs in TREX-DM,
  CERN-THESIS-2015-274, 2016 JINST TH 001.
  %%CITATION = CERN-THESIS-2015-274, 2016 JINST TH 001;%%

%\cite{Payez:2014xsa}
\bibitem{Payez:2014xsa}
  A.~Payez, C.~Evoli, T.~Fischer, M.~Giannotti, A.~Mirizzi and A.~Ringwald,
  Revisiting the SN1987A gamma-ray limit on ultralight axion-like particles,
  JCAP {\bf 1502}, 006 (2015).
  %doi:10.1088/1475-7516/2015/02/006
  %[arXiv:1410.3747 [astro-ph.HE]].
  %%CITATION = doi:10.1088/1475-7516/2015/02/006;%%
  %29 citations counted in INSPIRE as of 10 Oct 2016


%\cite{Meyer:2016wrm}
\bibitem{Meyer:2016wrm}
  M.~Meyer, M.~Giannotti, A.~Mirizzi, J.~Conrad and M.~A.~Sanchez-Conde,
  %``Fermi Large Area Telescope as a Galactic Supernovae Axionscope,''
  Phys.\ Rev.\ Lett.\  {\bf 118}, 011103 (2017)
  %doi:10.1103/PhysRevLett.118.011103
  %[arXiv:1609.02350 [astro-ph.HE]].
  %%CITATION = doi:10.1103/PhysRevLett.118.011103;%%
  %4 citations counted in INSPIRE as of 14 Mar 2017

%\cite{Ballou:2015cka}
\bibitem{Ballou:2015cka}
  R.~Ballou {\it et al.} (OSQAR Collaboration),
  New exclusion limits on scalar and pseudoscalar axionlike particles from light shining through a wall,
  Phys.\ Rev.\ D {\bf 92}, 092002 (2015).
  %doi:10.1103/PhysRevD.92.092002
  %[arXiv:1506.08082 [hep-ex]].
  %%CITATION = doi:10.1103/PhysRevD.92.092002;%%
  %9 citations counted in INSPIRE as of 22 Jan 2017

%\cite{DellaValle:2015xxa}
\bibitem{DellaValle:2015xxa}
  F.~Della Valle, A.~Ejlli, U.~Gastaldi, G.~Messineo, E.~Milotti, R.~Pengo, G.~Ruoso and G.~Zavattini,
  The PVLAS experiment: Measuring vacuum magnetic birefringence and dichroism with a
  birefringent Fabry–Perot cavity,
  Eur.\ Phys.\ J.\ C {\bf 76}, 24 (2016).
  %doi:10.1140/epjc/s10052-015-3869-8
  %[arXiv:1510.08052 [physics.optics]].
  %%CITATION = doi:10.1140/epjc/s10052-015-3869-8;%%
  %14 citations counted in INSPIRE as of 22 Jan 2017

%\cite{Li:2015tsa}
\bibitem{Li:2015tsa}
  D.~Li, R.~J.~Creswick, F.~T.~Avignone and Y.~Wang,
  Theoretical estimate of the sensitivity of the CUORE detector to solar axions,
  JCAP {\bf 1510}, 065 (2015).
  %doi:10.1088/1475-7516/2015/10/065
  %[arXiv:1507.00603 [astro-ph.CO]].
  %%CITATION = doi:10.1088/1475-7516/2015/10/065;%%
  %3 citations counted in INSPIRE as of 20 Jan 2017

%\cite{Archidiacono:2013cha}
\bibitem{Archidiacono:2013cha}
  M.~Archidiacono, S.~Hannestad, A.~Mirizzi, G.~Raffelt and Y.~Y.~Y.~Wong,
  Axion hot dark matter bounds after Planck,
  JCAP {\bf 1310}, 020 (2013).
  %[arXiv:1307.0615].
  %%CITATION = ARXIV:1307.0615;%%
  %11 citations counted in INSPIRE as of 09 Apr 2014

%\cite{Schlattl:1998fz}
\bibitem{Schlattl:1998fz}
  H.~Schlattl, A.~Weiss and G.~Raffelt,
  Helioseismological constraint on solar axion emission,
  Astropart.\ Phys.\  {\bf 10}, 353 (1999).
  %doi:10.1016/S0927-6505(98)00063-2
  %[hep-ph/9807476].
  %%CITATION = doi:10.1016/S0927-6505(98)00063-2;%%
  %74 citations counted in INSPIRE as of 20 Jan 2017

%\cite{Vinyoles:2015aba}
\bibitem{Vinyoles:2015aba}
  N.~Vinyoles, A.~Serenelli, F.~L.~Villante, S.~Basu, J.~Redondo and J.~Isern,
  New axion and hidden photon constraints from a solar data global fit,
  JCAP {\bf 1510}, 015 (2015).
  %doi:10.1088/1475-7516/2015/10/015
  %[arXiv:1501.01639 [astro-ph.SR]].
  %%CITATION = doi:10.1088/1475-7516/2015/10/015;%%
  %10 citations counted in INSPIRE as of 20 Jan 2017

%\cite{Bahre:2013ywa}
\bibitem{Bahre:2013ywa}
  R.~B\"ahre {\it et al.},
  Any Light Particle Search II --- Technical Design Report,
  JINST {\bf 8}, T09001 (2013).
  %doi:10.1088/1748-0221/8/09/T09001
  %[arXiv:1302.5647 [physics.ins-det]].
  %%CITATION = doi:10.1088/1748-0221/8/09/T09001;%%
  %91 citations counted in INSPIRE as of 10 Oct 2016

%\bibitem{Karuza:2015sia}
 % M.~Karuza, G.~Cantatore, A.~Gardikiotis, D.~H.~H.~Hoffmann, Y.~K.~Semertzidis and K.~Zioutas,
  %``KWISP: an ultra-sensitive force sensor for the Dark Energy sector,''
  %Phys.\ Dark Univ.\  {\bf 12}, 100 (2016).
  %doi:10.1016/j.dark.2016.02.004
  %[arXiv:1509.04499 [physics.ins-det]].
  %%CITATION = doi:10.1016/j.dark.2016.02.004;%%
  %1 citations counted in INSPIRE as of 04 Jan 2017.

%\cite{Krieger:2014wxa}
%\bibitem{Krieger:2014wxa}
 % C.~Krieger, K.~Desch, J.~Kaminski, M.~Lupberger and T.~Vafeiadis,
  %``An InGrid based Low Energy X-ray Detector for the CAST Experiment,''
  %PoS TIPP {\bf 2014}, 060 (2014).
  %%CITATION = POSCI,TIPP2014,060;%%
  %4 citations counted in INSPIRE as of 04 Jan 2017

\bibitem{spsc_report}
  K.~Desch (CAST collaboration), CAST Status Report to the SPSC for the 123rd Meeting, CERN-SPSC-2016-035.


\end{thebibliography}
\end{document}